\documentclass[ superscriptaddress, amsmath,amssymb, aps, showkeys]{revtex4-1}
\usepackage{graphicx}
\usepackage{dcolumn}
\usepackage{amsthm}
\usepackage{bm}

\pdfoutput=1
\usepackage{graphicx,graphics,epsfig,subfigure,times,bm,bbm,amssymb,amsmath,amsthm,mathrsfs,MnSymbol}
\usepackage{gensymb}
\usepackage{amsfonts}
\usepackage[matrix,frame,arrow]{xypic}
\usepackage[pdfstartview=FitH]{hyperref}
\hypersetup{
    colorlinks=true,       
    linkcolor=red,          
    citecolor=magenta,        
    filecolor=magenta,      
    urlcolor=cyan,           
    runcolor=cyan}
\usepackage{epstopdf}
\usepackage[pdftex]{color}
\usepackage{braket}
\usepackage{enumerate}
\usepackage[normalem]{ulem}
\usepackage[usenames,dvipsnames]{xcolor}
\usepackage{multirow}
\usepackage{mathtools}

\definecolor{orange}{rgb}{1,0.5,0}

\newcommand{\ignore}[1]{}
\usepackage{geometry}

\ignore{
\documentclass[eprintnumbers,amsmath,amssymb,onecolumn,a4paper,caption ]{article}
\usepackage{amsfonts}
\usepackage{amssymb}
\usepackage{mathrsfs}
\usepackage{mathbbold}
\usepackage{bbm}
\usepackage{mathrsfs}
\usepackage{dcolumn}
\usepackage{bm}
\usepackage{times,epsfig,amssymb,amsmath}
\usepackage{float}
\usepackage{subfigure}
\usepackage{geometry}\geometry{left=2.5cm,right=2.5cm,top=3cm,bottom=3cm}

\usepackage{color}

}

\begin{document}

\title{Diagonal Entropy and Topological Phase Transitions in Extended Kitaev Chains}

\author{Hong Qiao}
\affiliation{State Key Laboratory of Mesoscopic Physics, School of Physics, Peking University, Beijing 100871, China}

\author{Zheng-Hang Sun}
\email{zhenghangsun@outlook.com}
\affiliation{Institute of Physics, Chinese Academy of Sciences, Beijing 100190, China}
\affiliation{School of Physical Sciences, University of Chinese Academy of Sciences, Beijing 100190, China}

\author{Feng-Xiao Sun}
\affiliation{State Key Laboratory of Mesoscopic Physics, School of Physics, Peking University, Beijing 100871, China}
\affiliation{Collaborative Innovation Center of Quantum Matter, Beijing, China}

\author{Liang-Zhu Mu}
\affiliation{School of Physics, Peking University, Beijing 100871, China}

\author{Qiongyi He}
\email{qiongyihe@pku.edu.cn}
\affiliation{State Key Laboratory of Mesoscopic Physics, School of Physics, Peking University, Beijing 100871, China}
\affiliation{Collaborative Innovation Center of Quantum Matter, Beijing, China}
\affiliation{Collaborative Innovation Center of Extreme Optics, Shanxi University, Taiyuan 030006, China}
\affiliation{Beijing Academy of Quantum Information Sciences, Beijing 100193, China}

\author{Heng Fan}
\email{hfan@iphy.ac.cn}
\affiliation{Institute of Physics, Chinese Academy of Sciences, Beijing 100190, China}
\affiliation{School of Physical Sciences, University of Chinese Academy of Sciences, Beijing 100190, China}
\affiliation{Collaborative Innovation Center of Quantum Matter, Beijing, China}
\affiliation{CAS Central of Excellence in Topological Quantum Computation, Beijing 100190, China}
\affiliation{Beijing Academy of Quantum Information Sciences, Beijing 100193, China}

\begin{abstract}
We investigate the diagonal entropy for ground states of the extended Kitaev chains with extensive pairing and hopping terms.
The systems contain rich topological phases equivalently represented by topological invariant winding numbers and Majorana zero modes.
Both the finite size scaling law and block scaling law of the diagonal entropy are studied, which indicates that the diagonal entropy demonstrates volume effect. The parameter of volume term is regarded as the diagonal entropy density, which can identify the critical points of symmetry-protected topological phase transitions efficiently in the studied models, even for those with higher winding numbers.
The formulation of block scaling law and the capability of diagonal entropy density in detecting topological phase transitions are independent of the chosen bases. In order to manifest the advantage of diagonal entropy, we also calculate the global entanglement, which can not show clear signatures of the topological phase transitions.  This work provides a new quantum-informatic approach to characterize the feature of the topologically ordered states and may motivate a deep understanding of the quantum coherence and diagonal entropy in various condensed matter systems.
\end{abstract}
\maketitle

\section{Introduction}
Fractional quantum Hall state~\cite{Hall1} possesses exotic topological phase.
With the rapid development of condensed matter physics~\cite{c1,c2,c3} and topological quantum computation~\cite{t1,t2,t3,t4,MZM2},
topological phases and transitions, which are beyond the symmetry breaking theory and not characterized by local order parameter, become of great importance.
Symmetry-protected topological (SPT) phases and related phase transitions can be described by the topological order~\cite{XGWenBook} as well as the topological entanglement entropy\cite{TEE1,TEE2,TEE3,TEE4}. Over the past few years, one-dimensional topological systems, such as the extended Kitaev chains with variable-range pairing and hopping parameters in the Hamiltonian~\cite{model1,model2}, have received great attention. These models exhibit SPT phases and can produce new topological effects to enrich the appearance of one-dimensional topological superconductors~\cite{model3,model4} and have been demonstrated experimentally~\cite{exp1,exp2,exp3}, where Majorana zero modes (MZMs), as a key to realize fault-tolerant quantum computing~\cite{t1,t2,t3,t4,MZM2}, are observed.

It has been indicated that the entanglement entropy (EE) can provide information about the topological phases in extended Kitaev chains via the logarithmic scaling~\cite{model1,EElog1,EElog2,EElog4}. Moreover, quantum Fisher information (QFI), as a witness of multipartite entanglement~\cite{QFI},
can characterize topological quantum phases by a power-law scaling~\cite{QFI_Topo,QFI_Topo_YRZ}. Nevertheless,
characterizing the topological phases with high winding numbers $|\nu|\geq 2$ needs different approaches,
as shown by QFI defined on a dual lattice \cite{QFI_Topo_YRZ}. Additionally, in experiments, it is challenging to accurately measure the number of paired MZMs. Thus, developing new ways of characterizing SPT states with higher winding numbers from the perspective of quantum information is crucial.

On the other hand, quantum coherence is a useful resource for various quantum information tasks~\cite{QC1,QC2,QC3} and plays important roles in quantum critical systems~\cite{QC_QPT1,QC_QPT2,QC_QPT3,QC_QPT4,QC_QPT5} and quantum thermodynamics~\cite{DE,DE_a1,DE_a2}. Recently, quantum coherence has been quantified in a rigorous framework by the relative entropy~\cite{QC1}, $\mathcal{C}(\rho) = S(\rho^{\text{diag.}}) - S(\rho)$, where $S$ is the von Neumann entropy function. $S(\rho^{\text{diag.}})$ is the recently defined diagonal entropy (DE), for which $\rho^{\text{diag.}}$ denotes a diagonal part of the full density matrix $\rho$. In the case of a pure state $\rho$, the EE $S(\rho)=0$, and therefore the DE $S(\rho^{\text{diag.}})$ is equal to quantum coherence. Meanwhile, for pure states $S(\rho^{\text{diag.}})$ can be interpreted as the intrinsic randomness, which is proved to be a valid measure of quantum coherence~\cite{QC_XM}.
Based on above discussions, the DE of pure states can still provide information of the quantum coherence and thus may be possible to signal the quantum phases transitions.

More importantly, besides being detected in nanowire devices~\cite{exp1,exp2} and quantum spin liquids~\cite{exp3}, SPT states can also be realized by digital quantum simulation~\cite{exp4}, where the entanglement spectrum is applied to characterize the SPT states. However,
the calculation of both EE and entanglement spectrum require the reduced density matrix measured via state tomography in different bases, which is an obstacle for quantum simulation experiments. Fortunately, for DE, we only need to measure the diagonal part of a density matrix that can be obtained much easier than the full density matrix. Hence, DE is a more experimentally feasible quantity comparing with EE.

Recently, the global entanglement (GE), as a measure of multipartite entanglement~\cite{GE1,GE2}, was widely applied in the field of condensed matter physics.
The GE, which can be easily computed since only one-qubit reduced density operators are required,
is an indicator of quantum phase transitions~\cite{GE3,GE4,GE5} and many-body localization~\cite{GE6}. Nevertheless, whether the GE could detect TPTs remains an open question.

In this paper, we focus on the extended Kitaev chains with variable-range pairing and hopping and calculate the DE of the SPT
ground states. We also compute the GE in the models for reference.
The topological phases considered here are described by a $\mathbb{Z}$-valued topological invariant given by the winding number~\cite{WN1,WN2,WN3} as well as the pairs of MZMs~\cite{t1,t2,t3,t4,MZM2}. We show that the DE density extracted from both the finite size scaling law and block scaling law
can highlight the critical points associated with the topological phase transitions (TPTs) in the models for the TPTs, especially with higher winding numbers. However, the GE is trivial for TPTs in some cases.
It is worthwhile to emphasize that the characterization of TPTs by the DE density is independent of the chosen bases, showing the universality of this approach.

The remainder of this work is organized as follows. In Section 2, we briefly review the Hamiltonian of extended Kitaev chains, the winding numbers and MZMs in the models,
and the definition of DE and GE. In Section 3, the calculation results of diagonal entropy
as well as global entanglement in the extended Kitaev chains are presented. In Section 4, we report discussions and draw conclusions.

\section{Models and definitions}
\subsection{\label{sec:level2} Extended Kitaev chains}
First of all, we consider the extended Kitaev chains with variable-range pairing term, for which the Hamiltonian is written as~\cite{model1}
\begin{eqnarray}
\mathcal{H}_{1} = -\frac{J}{2}\sum_{j=1}^{N}(c_{j}^{\dagger}c_{j+1}+\text{H.c.})-\mu\sum_{j=1}^{N}(c_{j}^{\dagger}c_{j}-\frac{1}{2})
+ \frac{\Delta}{2}\sum_{j=1}^{N}\sum_{l=1}^{N-j}\frac{1}{d_{l}^{\alpha}}(c_{j}c_{j+l}+\text{H.c.}),
\label{hamiltonian1}
\end{eqnarray}
where $c_{j}^{\dagger}$ and $c_{j}$ are creation and annihilation spinless fermion operators, $J$ and $\Delta$ denote the strength of hopping and pairing, respectively. $\mu$ is the chemical potential. $\alpha$  (nonnegative)  represents the decay rate of pairing with distance, and $N$ is the size of the Kitaev chain. For closed boundary conditions, $d_{l}=\text{min}\{l,N-l\}$. It has been shown that the winding numbers of the topological phases in this model only contains $\nu = 0,\pm\frac{1}{2},\pm1$~\cite{model1,QFI_Topo}, however the phases of higher winding numbers are absent.

Next, in order to estimate the ability of DE to detect the TPTs with higher winding numbers, we also pay attention to the
extended Kitaev chain with both of longer-range pairing and hopping. The corresponding Hamiltonian reads~\cite{model2}
 \begin{eqnarray}
\mathcal{H}_{2} = -\mu\sum_{j=1}^{N}(c_{j}^{\dagger}c_{j}-\frac{1}{2})+ \sum_{l=1}^{r}\sum_{j=1}^{N-l}[\frac{\Delta}{d_{l}^{\alpha}}(c_{j}c_{j+l}+\text{H.c.})
-\frac{J}{d_{l}^{\beta}}(c_{j}^{\dagger}c_{j+l}+\text{H.c.})],
\label{hamiltonian2}
\end{eqnarray}
where $r$ represents the longest pairing and hopping distance, and $\beta$ (nonnegative) represents the decay rate of hopping.

Both of $\mathcal{H}_{1}$ and $\mathcal{H}_{2}$ are exactly solved by applying the Fourier transformation and Bogoliubov diagonalization \cite{ES1,ES2,ES3}. They can be written in momentum space with the form
 \begin{eqnarray}
\mathcal{H}=\sum_{k}\epsilon_{k}\Psi_{k}^{\dagger}(\textbf{h}_{k}\cdot\overrightarrow{\sigma})\Psi_{k},
\label{hk}
\end{eqnarray}
where the summation takes over all $k=(2\pi/N)(n+1/2)$ $(n=0,1,...,N-1)$ for closed chains with antiperiodic boundary conditions, $\epsilon_{k}$ is the energy spectra,   $\Psi_{k}^{\dagger}=(c_{k}^{\dagger},c_{-k})$ is the Nambu spinor, $\overrightarrow{\sigma}$ is the Pauli vector, and
 \begin{eqnarray}
\textbf{h}_{k}=(0,h_{y}(k),h_{z}(k))
\label{Av}
\end{eqnarray}
is the unit Anderson vector. The concrete expressions of the unit Anderson vectors for $\mathcal{H}_{1}$ and $\mathcal{H}_{2}$ are given in the Appendix A.

\subsection{\label{sec:level2} Winding numbers and Majorana zero modes}
Before we introduce the DE, we will present the definitions of winding numbers and Majorana zero modes (MZMs), as two characterizations of topologically nontrivial phases, and related results in the extended Kitaev chains. It has been proved that the number of pairs of MZMs is equal to the absolute value of winding number~\cite{QFI_Topo_YRZ}.

\begin{figure}[]
\includegraphics[width=0.8\textwidth]{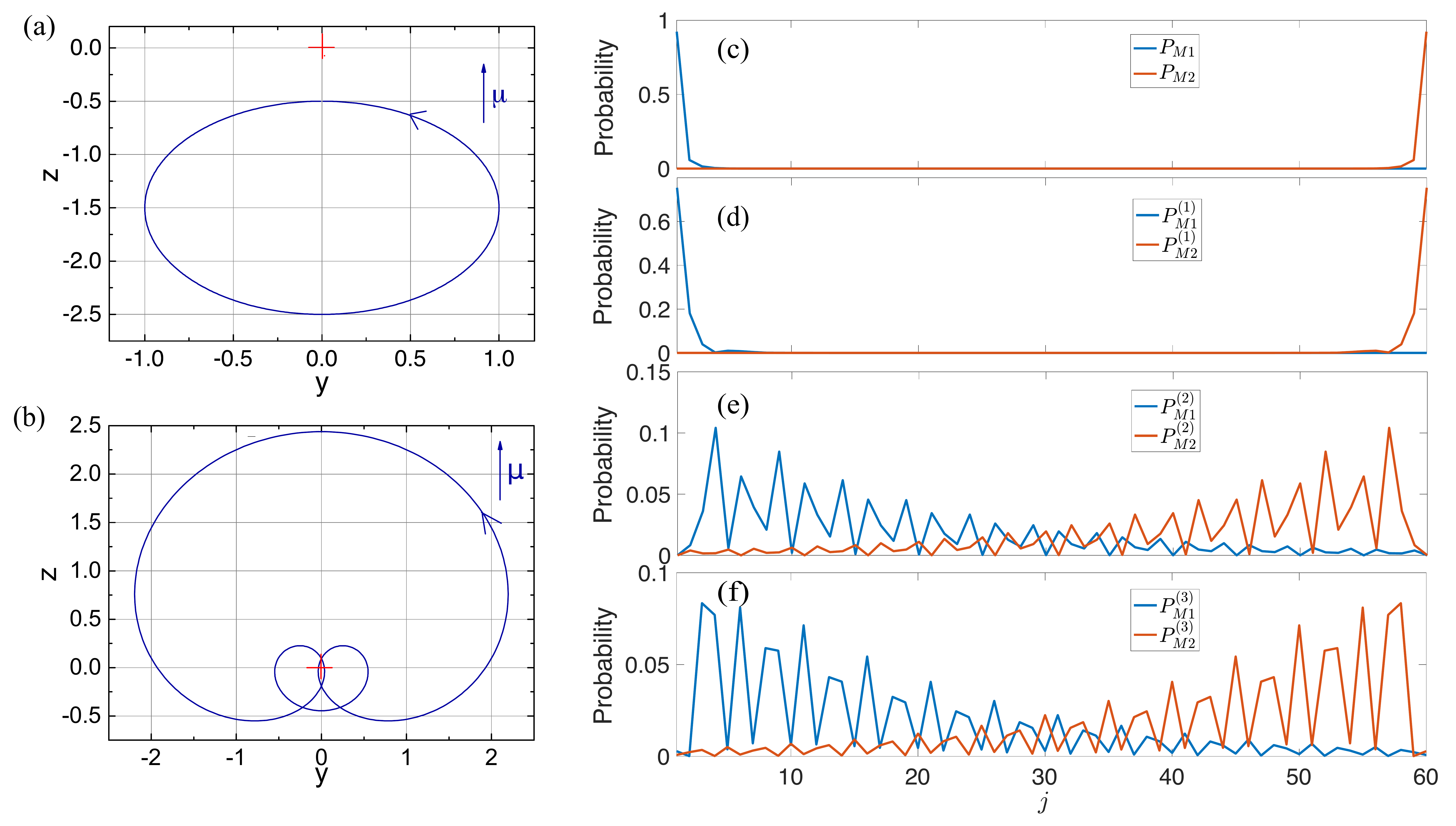}
\caption{(a) The trajectory of $\textbf{h}_{k}$ in $\mathcal{H}_{1}$ with $\alpha=+\infty$, $\Delta=-1$ and $\mu=-1.5$. In this case, the winding number $\nu=0$. (b) The trajectory of $\textbf{h}_{k}$ in $\mathcal{H}_{2}$ with $\alpha=\beta=0.2$, $r=3$, $J=-0.8$ and $\mu=-0.6$. In this case, the winding number $\nu=-3$. (c) For $\mathcal{H}_{1}$ with $\Delta=1$ and $\mu=-0.5$, we have only one pair of MZMs. For $\mathcal{H}_{2}$, when $\alpha=\beta=0.2$, $r=3$, $J=0.8$ and $\mu=0.6$, there are three pairs of MZMs shown in (d), (e) and (f).}\label{Fig_a3}
\end{figure}

The time reversal symmetry and particle-hole symmetry are preserved in both $\mathcal{H}_{1}$ and $\mathcal{H}_{2}$, which contributes to a constraint on the movement of the Anderson vector $\textbf{h}_{k}$ in the auxiliary $y-z$ plane as a circle $S^{1}$. The winding number defined as
 \begin{eqnarray}
\nu = \frac{1}{2\pi}\oint d \Theta = \frac{1}{2\pi}\int_{-\pi}^{\pi}dk \frac{1}{h_{y}(k)}\frac{\partial h_{z}(k)}{\partial k},
\label{wn}
\end{eqnarray}
where $\Theta_{k}=\arctan[h_{y}(k)/h_{z}(k)]/2$ is the Bogoliubov angle, can count how many times the vector $\textbf{h}_{k}$ winds in the $y-z$ plane when $k$ varies from $-\pi$ to $\pi$, i.e., the whole Brillouin zone. Hence, $\nu$ can characterize the mapping from the reduced Hamiltonian $\mathcal{H}_{k} = \textbf{h}_{k}\cdot\overrightarrow{\sigma}$ with $k \in S^{1}$ to the movement of $\textbf{h}_{k}$ and is regarded as a topological invariant. With the concrete expressions of $\textbf{h}_{k}$ given in Appendix A, we can directly obtain the trajectory of $\textbf{h}_{k}$. Two examples are presented in Fig. \ref{Fig_a3}. As shown in Fig. \ref{Fig_a3}(a), the trajectory shifts upward with the increase of $\mu$, and $\nu$ changes from 0 to -1 at $\mu=-1$ and from -1 to 0 at $\mu=1$. Similarly, as depicted in Fig. \ref{Fig_a3}(b), when $\mu$ increases, $\nu$ changes from $0$ to $-1$ at $\mu=-1.5$, from $-1$ to $-3$ at $\mu=-1$, and from $-3$ to $-1$ at $\mu\simeq-0.42$.

Next, in order to obtain the MZMs, we can rewrite $\mathcal{H}_{1}$ and $\mathcal{H}_{2}$ with open boundary conditions in terms of Majorana operators
 \begin{eqnarray}
a_{j}=\frac{1}{\sqrt{2}}(c_{j}^{\dagger}+c_{j}), b_{j}=\frac{i}{\sqrt{2}}(c_{j}^{\dagger}-c_{j}),
\label{MZM1}
\end{eqnarray}
with fermionic anticommutation relation $\{a_{i},a_{j}\}=\{b_{i},b_{j}\}=\delta_{ij}$ and $\{a_{i},b_{j}\}=0$.

We firstly consider a easier case, i.e., $\mathcal{H}_{1}$ with $\alpha=+\infty$, $\Delta=1$. It can be directly obtained that the trajectory of $\textbf{h}_{k}$ is the same as Fig. \ref{Fig_a3}(a) but with opposite direction and $\nu$ changes from 0 to 1 at $\mu=-1$ and from 1 to 0 at $\mu=1$ as $\mu$ increases. The Hamiltonian with Majorana operators reads
 \begin{eqnarray}
\mathcal{H}_{1}=\frac{i}{2}\sum_{j=1}^{N-1}b_{j}a_{j+1} - i\mu\sum_{j=1}^{N}a_{j}b_{j}.
\label{MZM2}
\end{eqnarray}
Based on the ansatz proposed in~\cite{model3,MZM1}, the left Majorana mode can be written as
 \begin{eqnarray}
\Phi_{\text{left}} = \sum_{j=1}^{L} m_{j}a_{j},
\label{MZM3}
\end{eqnarray}
where the real coefficients $m_{j}$ can be determined by the condition~\cite{MZM2,MZM3}
 \begin{eqnarray}
[\mathcal{H},\Phi_{\text{left}}]=0.
\label{MZM4}
\end{eqnarray}
Similarly, the right Majorana mode $\Phi_{\text{right}} = \sum_{j=1}^{L} n_{j}b_{j}$ with $n_{j}=m_{N-j+1}$ can also be obtained. In Fig. \ref{Fig_a3}(c), the probability of the left and right Majorana mode $P_{M1}$ and $P_{M2}$ for $\mu=-0.5$ are displayed. There is only one pair of MZMs, which is consistent with the winding number $\nu=1$.

We then study a more complex case. For the Hamiltonian $\mathcal{H}_{2}$, taking Eq. (\ref{MZM4}) into consideration, a straightforward calculation of $[\mathcal{H}_{2},\Phi_{\text{left}}]$ leads to
 \begin{eqnarray}
\sum_{l=1}^{r}\sum_{j=1}^{N-l}i[(\frac{\Delta}{d_{l}^{\alpha}}+\frac{J}{d_{l}^{\beta}})m_{j+l}b_{j} + (\frac{J}{d_{l}^{\alpha}}-\frac{\Delta}{d_{l}^{\beta}})m_{j}b_{j+l}] + i\mu\sum_{j=1}^{N}m_{j}d_{j} =0.
\label{MZM5}
\end{eqnarray}
For $\alpha=\beta=0.2$, $\Delta=1$, $J=0.8$ and $\mu=0.6$ (the trajectory of $\textbf{h}_{k}$ is the same as Fig. \ref{Fig_a3}(b) but with opposite direction and $\nu=3$), it can be found that there are only $k$ equations containing $k+3$ $m_{j}$. Consequently, there are 3 independent left zero modes, which  corresponds with the winding number $\nu=3$. The right Majorana modes can also be obtained using a similar approach. We depicted three pairs of MZMs in Fig. \ref{Fig_a3}(d)-(f). It can be observed that the signature of the second ($P_{M1}^{(2)}$, $P_{M2}^{(2)}$) and third ($P_{M1}^{(3)}$, $P_{M2}^{(3)}$) MZMs are weaker than the first MZMs, which may lead to a difficulty for experiments to characterize the number of MZMs.

\subsection{\label{sec:level2} Diagonal entropy}
Here, we focus on two types of scaling law for DE. (i) The finite size scaling law extracted from the association between the DE of pure state, that is, the total ground state without partial trace, and the system size $N$. (ii) The block scaling of DE obtained from the relation between the DE for a subsystem composed of $L$ continuous sites and $L$ in the considered systems with a fixed system size $N$ (usually $N\gg L$).

On the proposes of obtaining the first scaling law, we can calculate the DE of pure states in momentum space.
The ground state of $\mathcal{H}_{1}$ and $\mathcal{H}_{2}$ with momentum $k$ can be represented as
\begin{equation}
|g(k)\rangle = (\cos\Theta_{k} - i\sin\Theta_{k} \eta_{k}^{\dagger}\eta_{-k}^{\dagger})|0\rangle,
\label{GS}
\end{equation}
where $\eta_{k}$ and $\eta_{-k}^{\dagger}$ are Bogoliubov ferminic operators. $|0\rangle$ is the vacuum state.
Then, the DE can be calculated directly
\begin{eqnarray}
S_{k}^{\text{diag.}} = S(|g(k)\rangle\langle g(k)|)
= -\sin^{2}\Theta_{k}\log(\sin^{2}\Theta_{k})-\cos^{2}\Theta_{k}\log(\cos^{2}\Theta_{k}),
\label{DE_k}
\end{eqnarray}
where $S(\rho)=-\text{Tr}(\rho\log\rho)$, and the DE of the $N$ sites ground state is $S^{\text{diag.}}_{N}=\sum_{k}S_{k}^{\text{diag.}}$ since the ground state is $|\mathcal{G}\rangle=\prod_{k} |g(k)\rangle$.
It is noted that we pay attention to the DE of pure state, that is, the total ground state without partial trace.
Therefore, DE has the potential to capture nonlocal properties of the ground states and detect TPTs.

We also study the second scaling law of DE in the extended Kitaev chains with infinite length to ensure $N\gg L$. Before the derivation of the diagonal reduced density matrix $\rho_{L}^{\text{diag.}}$, for better comprehension of the block scaling law of DE, we emphasize that the extended Kitaev chains $\mathcal{H}_{1}$ and $\mathcal{H}_{1}$ can be mapped to extended Ising models $\mathcal{H}_{1}^{\text{Ising}}$ and $\mathcal{H}_{2}^{\text{Ising}}$ respectively via the Jordan-Wigner transformation. $\mathcal{H}_{1}^{\text{Ising}}$ reads
\begin{equation}
\mathcal{H}_{1}^{\text{Ising}} = \sum_{j=1}^{N}\sum_{l=1}^{N-j} (\frac{J_{l}^{x}}{2}\sigma_{j}^{x}\sigma_{j+l}^{x} +\frac{J_{l}^{y}}{2}\sigma_{j}^{y}\sigma_{j+l}^{y})\prod_{n=j+1}^{j+l-1}\sigma_{n}^{z} + \sum_{j=1}^{N}\frac{\mu}{2} \sigma_{j}^{z},
\label{Ising_1}
\end{equation}
where $J_{l}^{x} = -\frac{1}{2}(\frac{J}{2}-\frac{\Delta}{d_{l}^{\alpha}})$, $J_{l}^{y} = -\frac{1}{2}(\frac{J}{2}-\frac{\Delta}{d_{l}^{\alpha}})$ for $l=1$ and $J_{l}^{x} = -J_{l}^{y} = \frac{\Delta}{2 d_{l}^{\alpha}}$ for other values of $l$, and $\mathcal{H}_{2}^{\text{Ising}}$ is given by
\begin{equation}
\mathcal{H}_{2}^{\text{Ising}} = \sum_{l=1}^{r}\sum_{j=1}^{N-l} (\frac{J_{l}^{x}}{2}\sigma_{j}^{x}\sigma_{j+l}^{x} +\frac{J_{l}^{y}}{2}\sigma_{j}^{y}\sigma_{j+l}^{y})\prod_{n=j+1}^{j+l-1}\sigma_{n}^{z} + \sum_{j=1}^{N}\frac{\mu}{2} \sigma_{j}^{z},
\label{Ising_2}
\end{equation}
where $J_{l}^{x} = -\frac{1}{2}(\frac{J}{d_{l}^{\beta}}+\frac{\Delta}{d_{l}^{\alpha}})$, $J_{l}^{y} = -\frac{1}{2}(\frac{J}{d_{l}^{\beta}}-\frac{\Delta}{d_{l}^{\alpha}})$ for all values of $l$.

Then the diagonal terms of the reduced density matrix $\rho_{L}^{\text{diag.}}$ for ground state can be always expended as
\begin{equation}
\rho^{\text{diag.}}_{L} = \frac{1}{2^{L}}\sum_{\substack{\gamma_{1},..., \gamma_{L},\in\{0,z\}}}\langle\sigma^{\gamma_{1}}_{1}...\sigma^{\gamma_{L}}_{L}\rangle\sigma^{\gamma_{1}}_{1}...\sigma^{\gamma_{L}}_{L}, \label{diagonal_matrix}
\end{equation}
where $\sigma^{z}_{l}$ and $\sigma^{0}_{l}$ $(l=1,...,L)$ are the Pauli matrix and two-dimensional identity matrix, respectively. The correlation functions $\langle\sigma^{\gamma_{1}}_{1}...\sigma^{\gamma_{L}}_{L}\rangle$ can be calculated by applying the Wick theorem, which is shown in Appendix B.
Then the diagonal entropy $S(\rho_{L}^{\text{diag.}}) = -\text{Tr}(\rho_{L}^{\text{diag.}}\log\rho_{L}^{\text{diag.}})$ can be obtained.

\subsection{\label{sec:level2} Global entanglement}
Next, we present the definition of GE, which is give by~\cite{GE1,GE2}
\begin{equation}
E = \frac{2}{N}\sum_{i=1}^{N}[1-\text{Tr}(\rho_{i}^{2})], \label{GE}
\end{equation}
where $\rho_{i}$ denotes the one-qubit reduced density operator. In the extended Ising chains (\ref{Ising_1}) and (\ref{Ising_2}),
$\rho_{i}$ can be written as~\cite{single_q1,single_q2}
\begin{equation}
\rho_{i} = \frac{1}{2}\left(
                        \begin{array}{cc}
                          1+\langle\sigma_{i}^{z}\rangle & 0 \\
                          0 & 1-\langle\sigma_{i}^{z}\rangle \\
                        \end{array}
                      \right)
, \label{single_q}
\end{equation}
with $\langle\sigma_{i}^{z}\rangle=\frac{1}{2\pi}\int_{-\pi}^{\pi}e^{-2i\Theta_{k}}dk$ (for the chains with infinite length). The expressions of Bogoliubov angle $\Theta_{k}$ are presented in Appendix A.

\section{Results}
\begin{figure}
\includegraphics[width=0.6\textwidth]{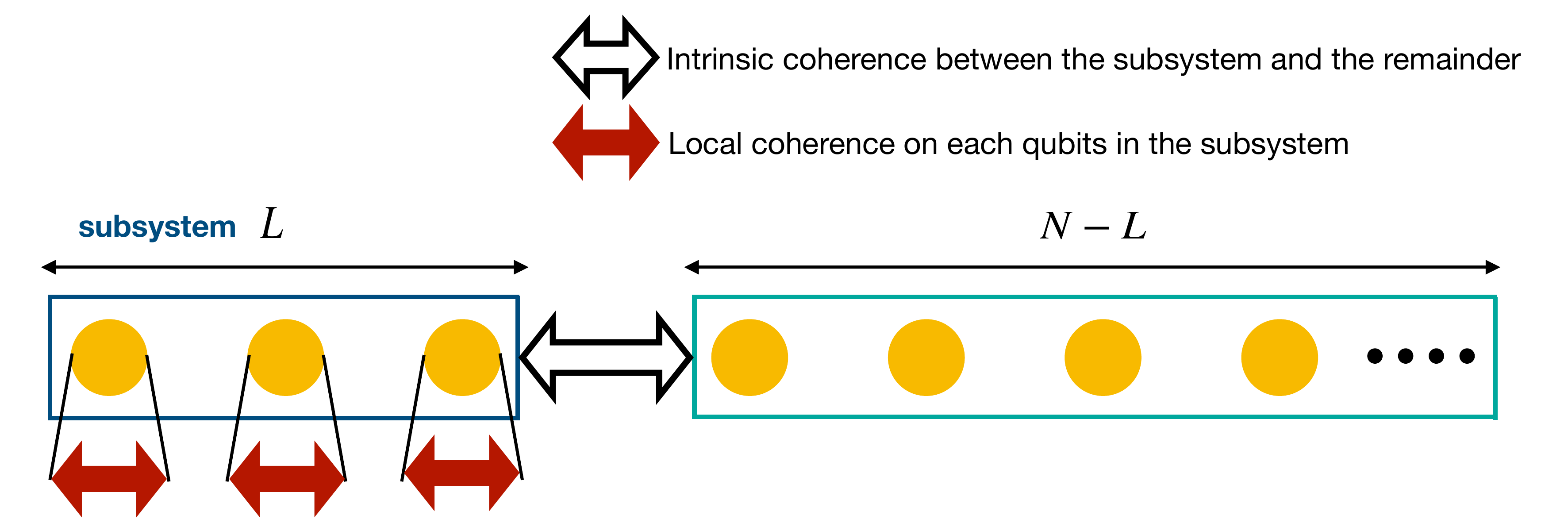}
\caption{A diagrammatic sketch of the distribution of quantum coherence. The quantum coherence of the subsystem consisted of $L$ continuous spins $\rho_{L}$ is $\mathcal{C}(\rho_{L}) = S(\rho_{L}^{\text{diag.}}) - S(\rho_{L})$. There are two contributions of $\mathcal{C}(\rho_{L})$, the intrinsic coherence and local coherence, which measure the coherence between the bipartite system and the coherence on each qubits in the considered subsystem respectively. The intrinsic coherence the local coherence on each qubits are marked by the arrows. The total local coherence is directly proportional to the number of spins $L$ in the subsystem.}\label{Fig0}
\end{figure}

\begin{figure}
\includegraphics[width=0.6\textwidth]{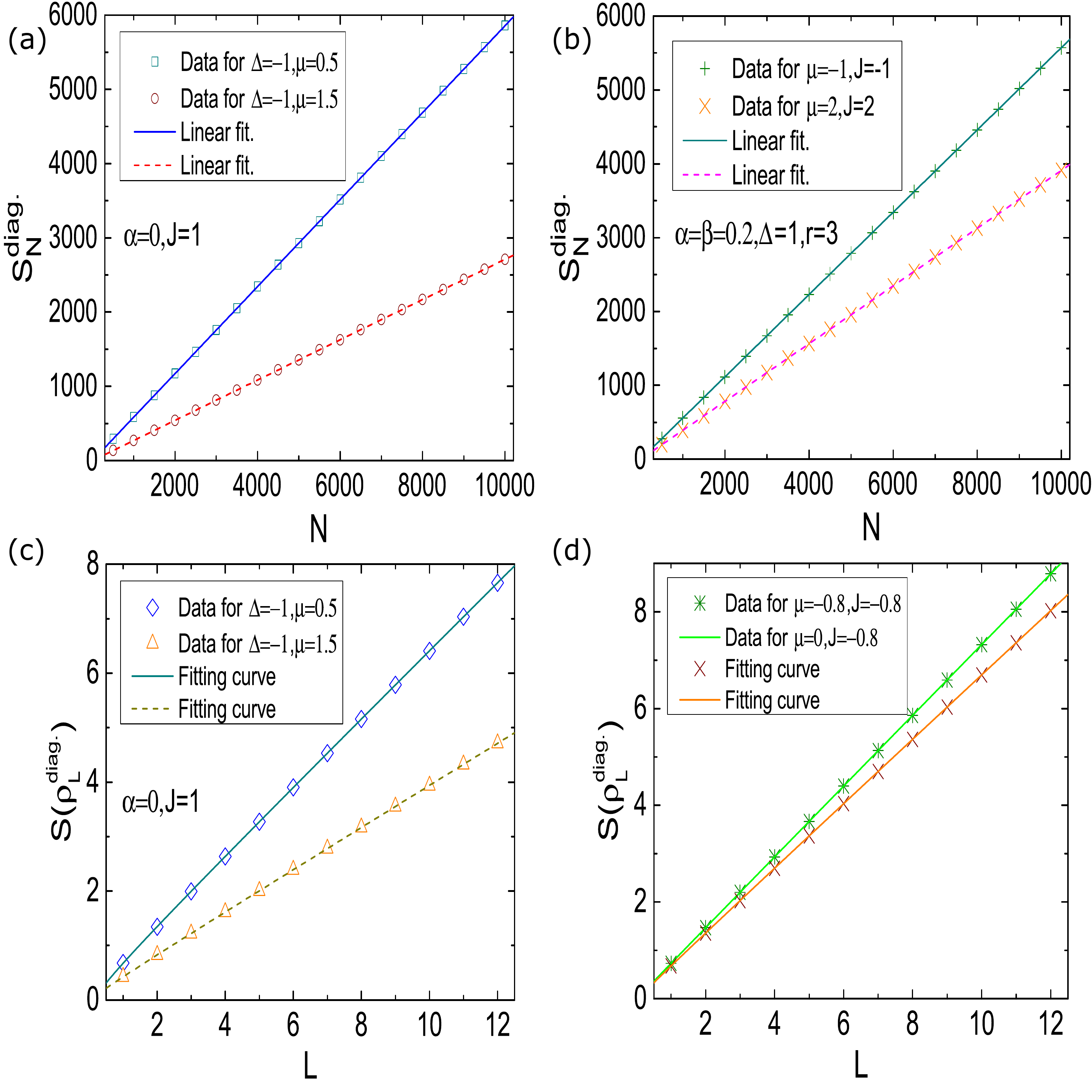}
\caption{The DE of pure states $S_{N}^{\text{diag.}}$ as a function of the length of chains $N$ for $\mathcal{H}_{1}$ (a) and $\mathcal{H}_{2}$ (b). The DE
for a subsystem composed of $L$ continuous sites $S(\rho_{L}^{\text{diag.}})$ as a function of the length of subsystem $L$ for $\mathcal{H}_{1}$ (c) and $\mathcal{H}_{2}$ (d).}\label{Fig1}
\end{figure}
\subsection{\label{sec:level2} Scaling laws of diagonal entropy}
We firstly explore the formulation of finite size scaling law of DE. The size of extended Kitaev chains reads $N$.
As shown in Figs. \ref{Fig1}(a) and \ref{Fig1}(b), the numerical calculations indicate that there is a volume law for DE, i.e.,
\begin{equation}
S^{\text{diag.}}_{N}=sN. \label{DE_scaling_k}
\end{equation}
It has been recognized that the validity of scaling laws may depend on the winding numbers of topological phases. For instance, the power-law scaling of QFI ~\cite{QFI_Topo_YRZ} and two-site scaling law of quantum coherence~\cite{QC_QPT1} for the topological phases with high winding numbers can only be observed in the dual lattice. It is remarkable that the scaling law (\ref{DE_scaling_k}) is not only true for the lower $\nu$ topological phases shown in Fig.~\ref{Fig1}(a) but also valid for the high $\nu=\pm 3$ topological phases indicated in Fig.~\ref{Fig1}(b).

Then, we study the block scaling law of DE for the ground states of the extended Kitaev chains.
The block scaling law show the relation between some quantities, such as EE, and the length of
subsystem $L$ in a system with size $N$. The block scaling law of EE for the ground states of many-body systems,
as the area law of EE~\cite{TEE1,TEE2,TEE3,EElog3,summary_EE1,summary_EE2}, is fully explored. Moreover,
the volume law of the R\'{e}nyi entropy for scrambled pure quantum states~\cite{volume_1}
and the DE for the ground states of quantum magnetism systems~\cite{volume_2} are revealed.

The results presented in Fig. \ref{Fig1}(c) and \ref{Fig1}(d) indicate that the block scaling law of DE is
\begin{equation}
S(\rho_{L}^{\text{diag.}})=aL+b\log L+c, \label{scaling_block}
\end{equation}
which is still satisfied for the topological phases with higher winding number. The logarithmic term originates from EE because $S(\rho_{L}^{\text{diag.}}) = \mathcal{C}(\rho_{L}) + S(\rho_{L})$, and the block scaling law of EE for reduced ground states is $S(\rho_{L})\sim \log L$~\cite{EElog3}.
The volume term can be explained from the perspective of distribution of quantum coherence~\cite{QC_QPT1,basis1}. The total quantum coherence can be divided into the intrinsic coherence and the local coherence, which measure the coherence between different parts and on each qubits respectively, see Fig. \ref{Fig0} for details. Thus, the local coherence, as a contribution of diagonal entropy, is naturally directly proportional to the number of qubits (the length of subsystem $L$). We also note that
only the volume term survives for the DE of pure states since both the EE and intrinsic coherence vanish for the total ground state without partial trace.
\begin{figure*}[t]
	\centering
	\includegraphics[width=0.9\textwidth]{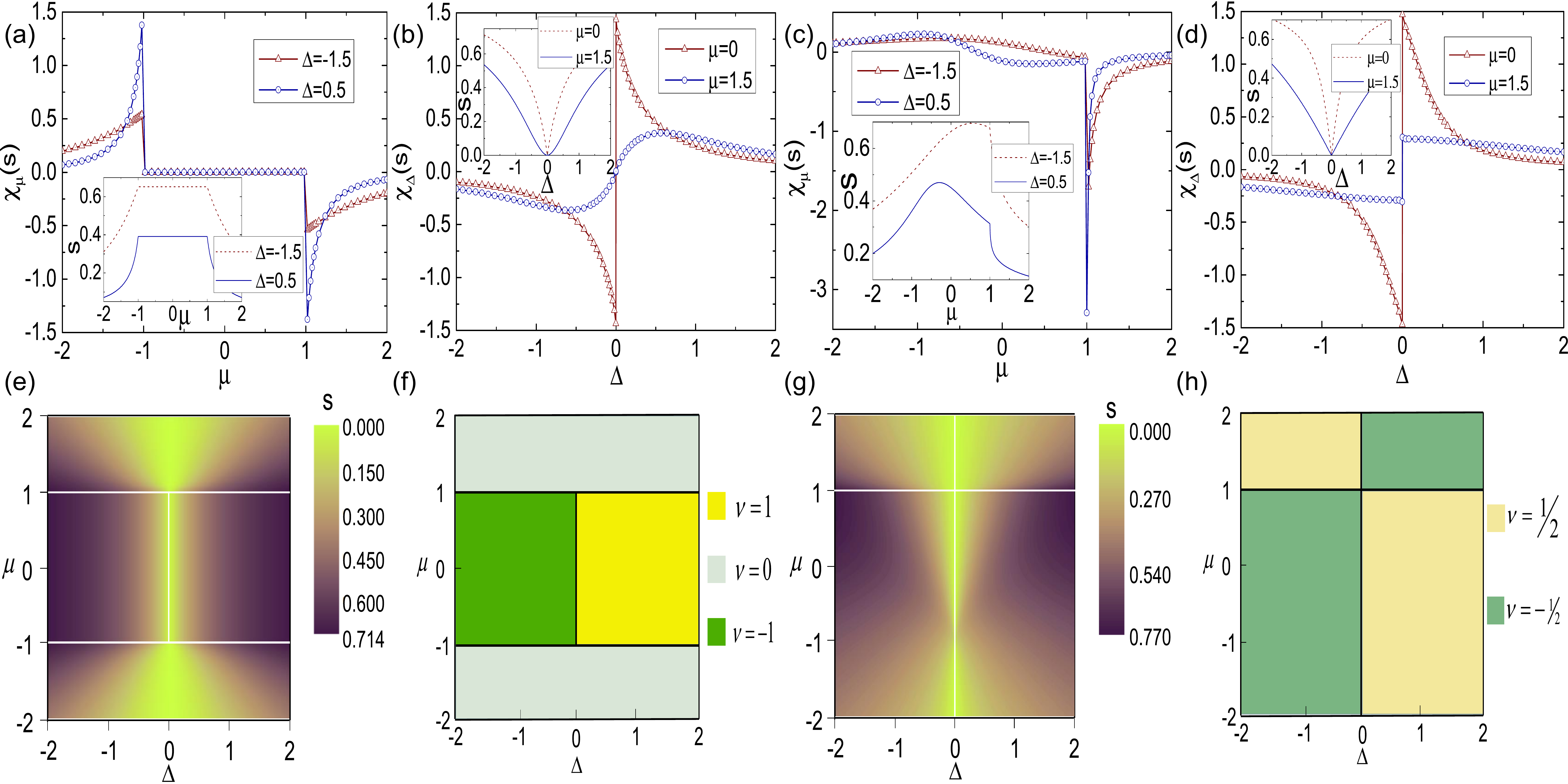}
	\caption{The susceptibility of DE density $\chi_{\mu}(s)$ as a function of $\mu$ with $\Delta=-1.5,0.5$ (a) and as a function of  $\Delta$ with $\mu=0,1.5$ (b). The insets show the $s$ as a function of $\mu$ and $\Delta$, respectively. (e) The value of $s$ in the $\mu-\Delta$ plane, where the locations of non-analyticality of $s$ are marked by the solid white lines. (f) Phase diagram of $\mathcal{H}_{1}$ in the $\mu-\Delta$ plane, where topological phases are identified by different winding numbers $\nu$. Four cases are for the extended Kitaev chain $\mathcal{H}_{1}$ with $\alpha=+\infty$ and $J=1$. (c) (d) (g) (h) are similar to (a) (b) (e) (f) but with the parameters $\alpha=0$. It should be noted in (b) that $s$ is analytical and therefore the DE susceptibility $\chi_{\Delta}(s)$ is continuous at $\Delta=0$ with parameters $\alpha=+\infty$, $J=1$, and $\mu=1.5$, which suggests that the TPT doesn't occur under this condition.
	}\label{Fig2}
\end{figure*}
\subsection{\label{sec:level2} Results of diagonal entropy density}
In the following, we pay attention to the DE density defined as $s=S^{\text{diag.}}_{N}/N$ for the finite size scaling law of DE and as the parameter $a$ in Eq. (\ref{scaling_block}) for the block scaling law of DE. We then explore its capability in signaling the existence of TPTs in the studied model.
We calculate the DE density $s$ in the extended Kitaev chain described by $\mathcal{H}_{1}$. As shown in the insets of Figs.~\ref{Fig2}(a-d), $s$ behaves non-analytically at critical points where TPTs occur. In order to probe the TPTs, the idea of susceptibility is employed~\cite{QC_QPT2}. The DE density susceptibility with respect to the quantity $\mathcal{O}$ which drives TPTs is defined as $\chi_{\mathcal{O}}(s)=\partial s/ \partial\mathcal{O}$. The non-analytical behaviors of $s$ at critical points lead to the discontinuity of DE susceptibility $\chi_{\mathcal{O}}(s)$. From Figs.~\ref{Fig2}(a-d), we can find that the DE susceptibility $\chi_{\mu}(s)$ is able to spotlight the critical points of TPTs efficiently.
The values of $s$ in the $\mu-\Delta$ plane with $\alpha=+\infty$ and $\alpha=0$ are shown in Figs. \ref{Fig2}(e) and \ref{Fig2}(g) respectively. The locations of non-analytical DE are highlighted by the solid white lines, which are consistent with the phase diagrams shown in Figs.~\ref{Fig2}(f) and \ref{Fig2}(h) obtained by numerically calculating the winding numbers.
\begin{figure}
\includegraphics[width=0.6\textwidth]{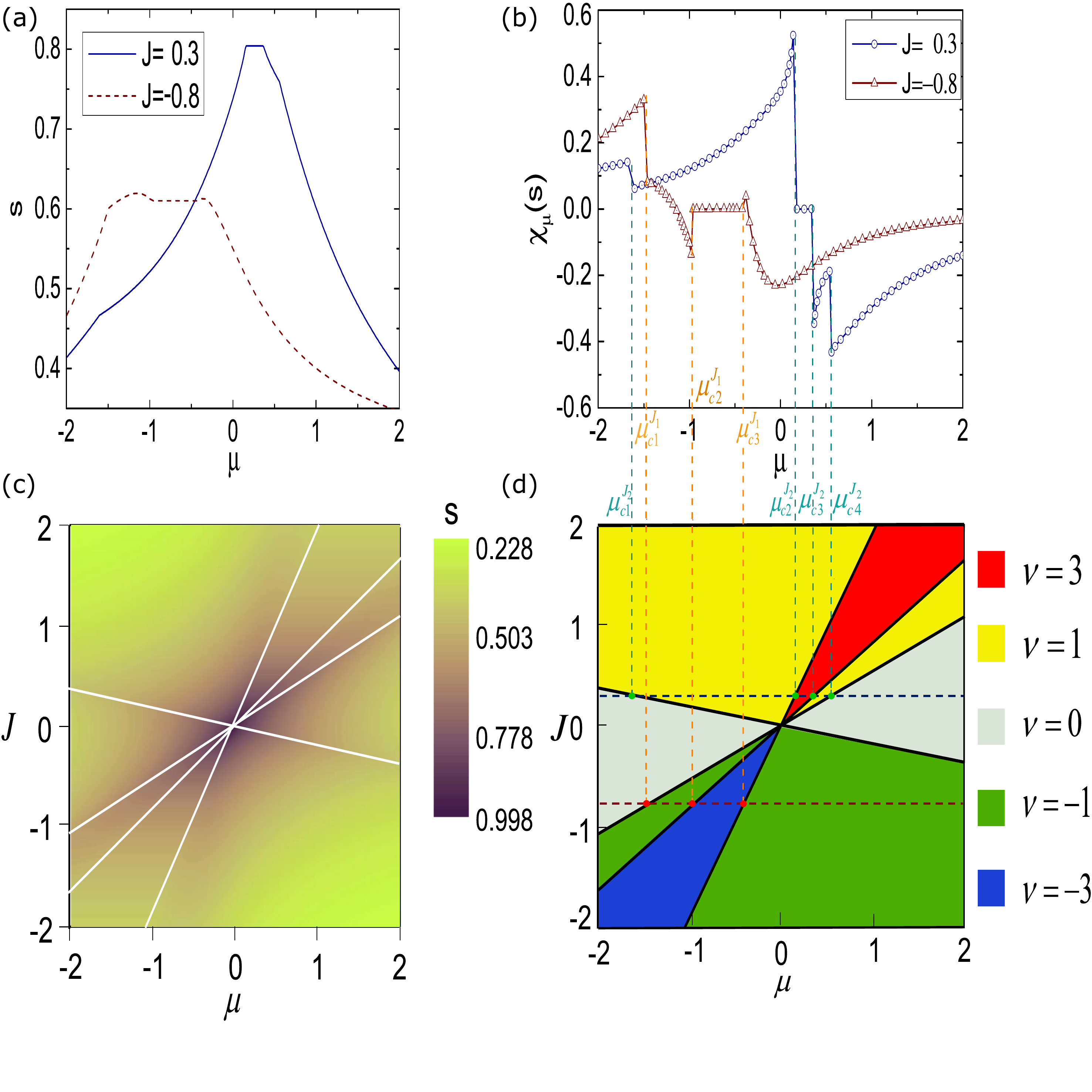}
\caption{(a) The DE density $s$ and (b) the DE susceptibility $\chi_{\mu}(s)$ as functions of $\mu$ with $r=3$, $\alpha=\beta=0.2$, $\Delta=1$ and $J=0.3,-0.8$ in the model $\mathcal{H}_{2}$. (c) The values of $s$ in the $\mu-\Delta$ plane with $\alpha=\beta=0.2$ and $\Delta=1$. The solid white lines mark the locations of discontinuous $\chi_{\mu}(s)$. (d) Topological phases in $\mathcal{H}_{2}$ with $\alpha=\beta=0.2$ and $\Delta=1$ characterized via winding numbers. It is noted that for $J_{1}=-0.8$, the locations of the critical points are $\mu_{c1}^{J_{1}}\simeq -1.5$, $\mu_{c2}^{J_{1}}\simeq -1$, $\mu_{c3}^{J_{1}}\simeq -0.42$, which are highlighted by the dashed red lines in (d), and for $J_{2}=0.3$, the locations of the critical points are $\mu_{c1}^{J_{2}}\simeq -1.6$, $\mu_{c2}^{J_{2}}\simeq 0.15$, $\mu_{c3}^{J_{2}}\simeq 0.36$, $\mu_{c4}^{J_{2}}\simeq 0.55$, which are highlighted by the dashed blue lines in (d). The locations of the critical points are consistent with the locations of discontinuous points as shown in (b).}\label{Fig3}
\end{figure}

In order to demonstrate the capability of DE in detecting TPTs related to the topological phases with high winding numbers, we study the DE in the extended Kitaev chain with both longer-range pairing and hopping described by $\mathcal{H}_{2}$. The phase diagram of $\mathcal{H}_{2}$ with $r=3$, $\alpha=\beta=0.2$, and $\Delta=1$ is shown in Fig.~\ref{Fig3}(d), and we can see that there exist the topological phases with winding numbers $|\nu_{\text{max}}|=3$ in this model.
The dependence of values of DE density $s$ and its susceptibility $\chi_{\mu}(s)$ on parameter $\mu$ are displayed in Figs.~\ref{Fig3}(a) and ~\ref{Fig3}(b). The solid white lines in Fig.~\ref{Fig3}(c) mark the locations of discontinuous DE susceptibility, which coincide with the phase diagram of $\mathcal{H}_{2}$ given in Fig.~\ref{Fig3}(d). We can see that all the properties of the symmetry-protected TPTs including higher winding numbers can be sufficiently captured by the DE susceptibility.

\begin{figure}
\includegraphics[width=0.6\textwidth]{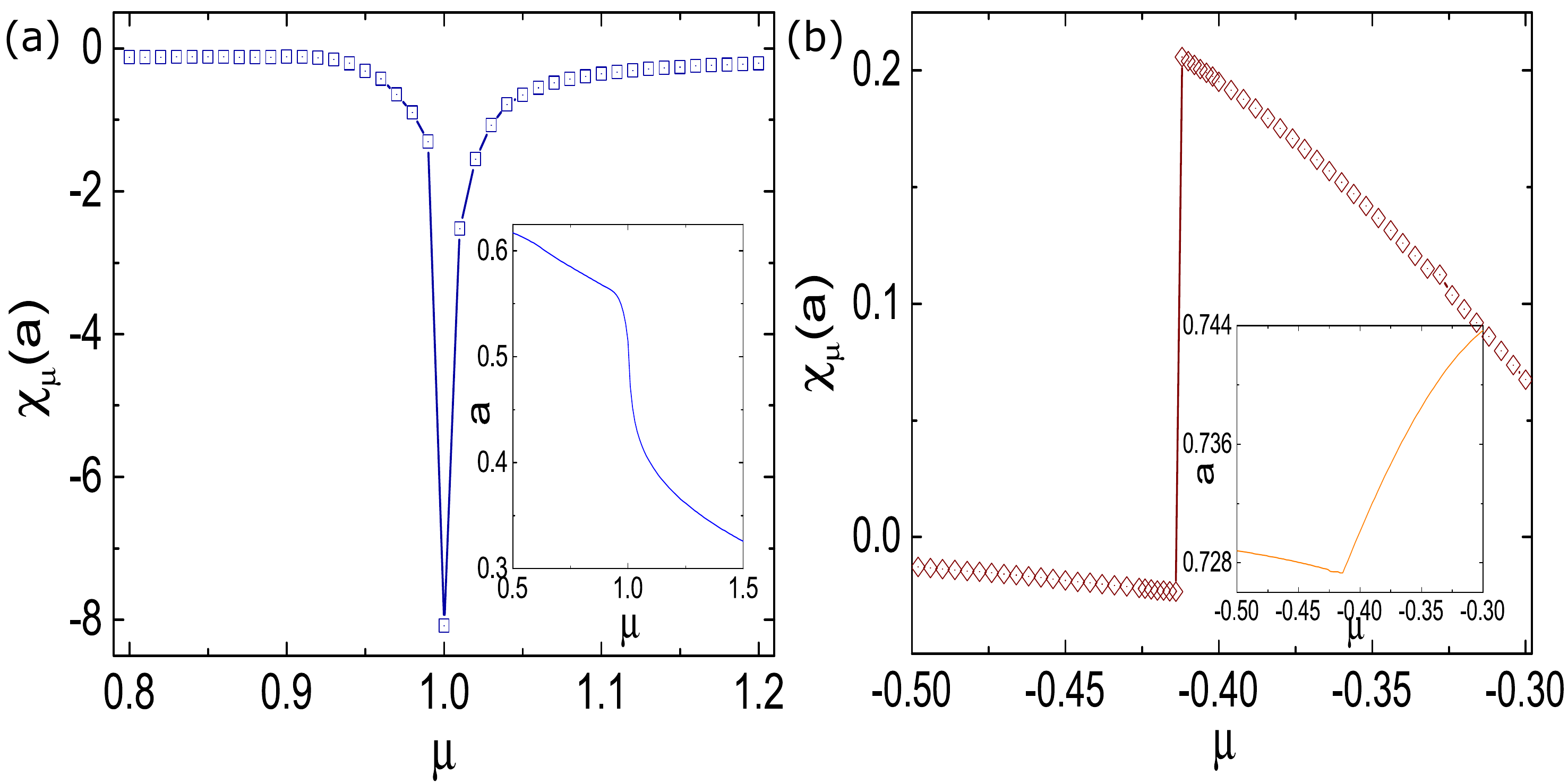}
\caption{The DE density $a$ for the block scaling law (inset) and the susceptibility of $a$ $\chi_{\mu}(a)$  as a function of $\mu$ for $\mathcal{H}_{1}$ with $\Delta = -1$, $\alpha=0$, $J=1$ (a) and for $\mathcal{H}_{2}$ with $\Delta = 1$, $\alpha=\beta=0.2$, $J=-0.8$ (b). As shown in Figs. \ref{Fig2}(c) and \ref{Fig3}(c), the locations of TPTs for the parameters in (a) and (b) are $\mu_{c}=1$ and $\mu_{c}\simeq -0.42$ respectively. }\label{Fig4}
\end{figure}

Furthermore, we plot the DE density extracted from the block scaling law $S(\rho_{L}^{\text{diag.}})=aL+b\log L+c$, i.e., the parameter $a$, as a function of $\mu$ for $\mathcal{H}_{1}$ with $\Delta = -1$, $\alpha=0$, $J=1$ in Fig. \ref{Fig4}(a). Comparing to the results of $\chi_{\mu}(s)$ at $\mu=1$ shown in Fig. \ref{Fig2}(c), the non-analytical behaviors of $\chi_{\mu}(a)$ at $\mu=1$ is similar to the DE density obtained from finite size scaling law.
Thus the TPTs can be diagnosed via the discontinuity of susceptibility $\chi_{\mu}(a) = \partial{a}/\partial{\mu}$.
Meanwhile, we can study the DE density $a$ for $\mathcal{H}_{2}$ with $\Delta = 1$, $\alpha=\beta=0.2$, where the topological phase with winding number $\nu=-3$ exists. Remarkably, as depicted in Fig. \ref{Fig4}(b), the DE density $a$ can also characterize the TPTs with high winding number topological phases. In comparison with~\cite{volume_2}, where the DE density $a$ in Ising model is studied, the non-analytical behaviors of $a$ can not be observed at the critical point associated with the quantum phase transition between ferromagnetic phase and paramagnetic phase. Instead, there is a dramatic change of $a$ and $a$ remains analytical and continuous at the critical point, which makes a distinction between the capability of $a$ in detecting (second-order) quantum phase transitions and topological phase transitions.  

\begin{figure}
\includegraphics[width=0.7\textwidth]{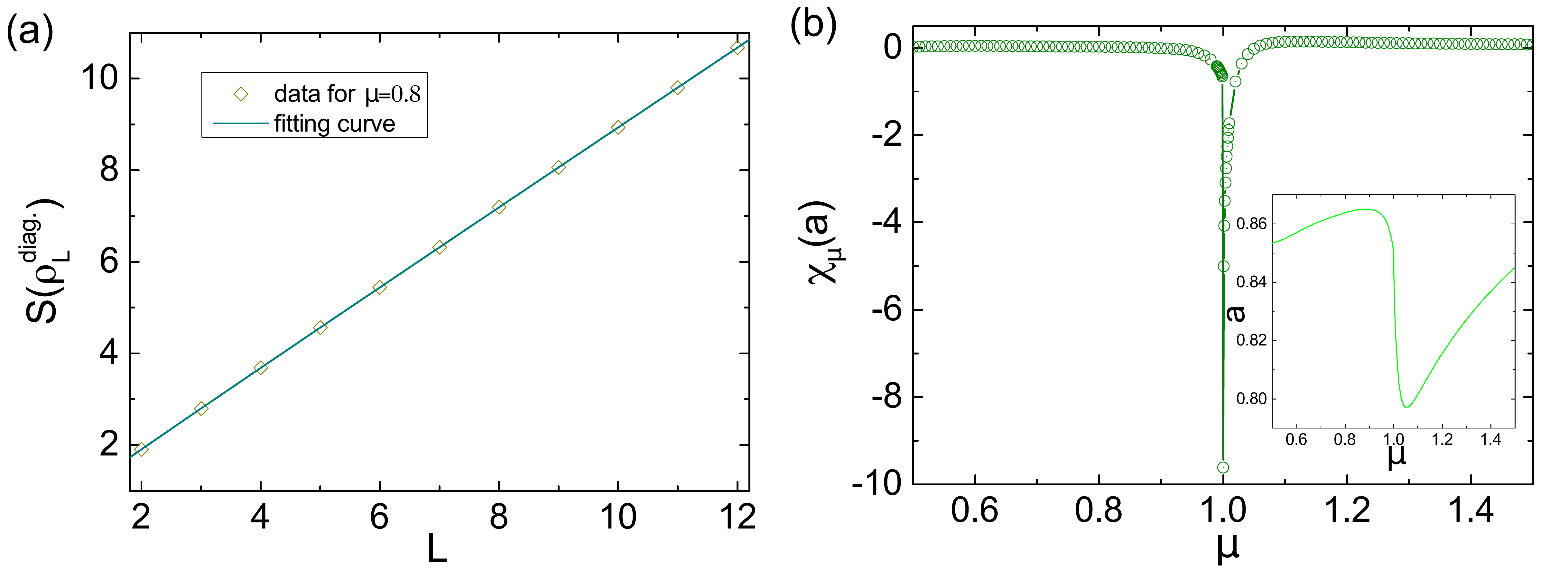}
\caption{(a) The value of DE $S(\rho_{L}^{\text{diag}.})$ as a function of subsystem size $L$ for $\mathcal{H}_{1}$ with $\Delta=-1$, $\alpha=0$, $J=1$ and $\mu=0.8$. The form of fitting curve satisfies Eq.~(\ref{DE}). (b) The parameter $a$ and its susceptibility with respect of $\mu$ $\chi_{\mu}(a)$ as a function of $\mu$ for $\mathcal{H}_{1}$ with $\Delta=-1$, $\alpha=0$, $J=1$.}\label{Fig_a}
\end{figure}

Finally, we recognize that the DE is a basis-dependent quantity~\cite{basis1,basis2}. In above discussions, we only consider the DE in $\sigma^{z}$ basis. In order to demonstrate whether the formulation of DE block scaling law Eq.~(\ref{scaling_block}) and the capability of DE in detecting TPTs are dependent on the chosen basis or not, we numerically study the DE in $\sigma_{x}$ basis. As depicted in Fig. \ref{Fig_a} (a), the formulation of block scaling law for DE in $\sigma_{x}$ basis still satisfies the form of Eq.~(\ref{scaling_block}). Therefore, we can obtain the DE density $a$ by fitting the data of DE as a function of subsystem size $L$. The parameter $a$ and its susceptibility with respect of $\mu$, i.e., $\chi_{\mu}(a)$, as a function of $\mu$ are shown in Fig. \ref{Fig_a},
which indicates that the DE in $\sigma_{x}$ basis can be a probe of the critical points associated with TPTs. Hence, we argue that the formulation of DE block scaling law as well as the capability of DE in characterizing TPTs are independent of the chosen basis. As a side remark, in Appendix B, we study the parameters $b$ and $c$ in Eq. (\ref{scaling_block}), which suggests that the parameters $b$ and $c$ can also diagnose the TPTs.

\subsection{\label{sec:level2} Results of global entanglement}
\begin{figure}
\includegraphics[width=1\textwidth]{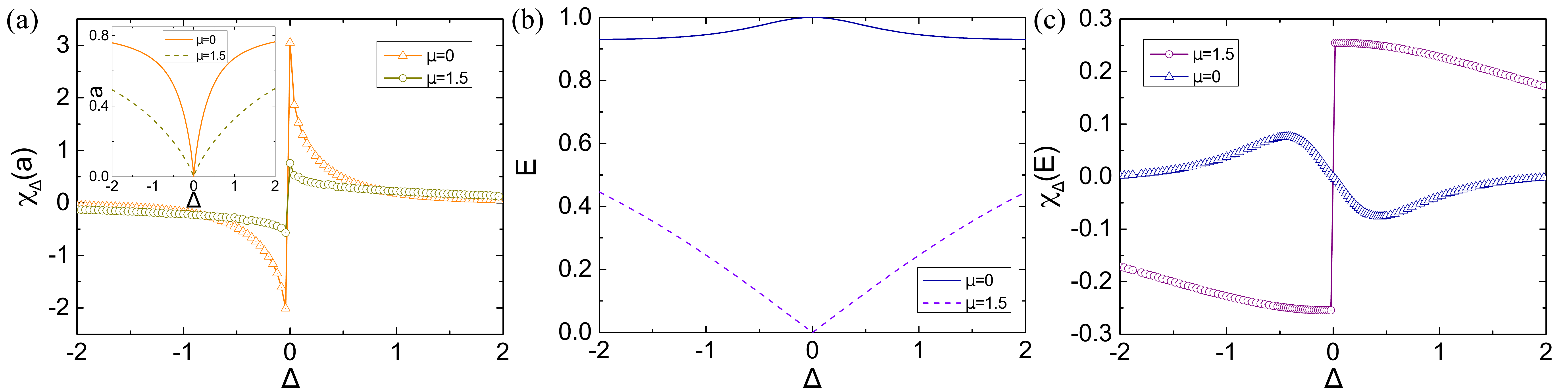}
\caption{(a) The susceptibility of DE density $\chi_{\Delta}(a)$ as a function of $\Delta$ in $\mathcal{H}_{1}$ with $\alpha=0$ and $\mu=0,1.5$. The inset shows the DE density $a$ as a function of $\Delta$ in $\mathcal{H}_{1}$ with $\alpha=0$ and $\mu=0,1.5$. (b) The GE $E$ as a function of $\Delta$ in $\mathcal{H}_{1}$ with $\alpha=0$ and $\mu=0,1.5$. (c) The susceptibility of GE $\chi_{\Delta}(E)$ as a function of $\Delta$ in $\mathcal{H}_{1}$ with $\alpha=0$ and $\mu=0,1.5$.}\label{Fig_a4}
\end{figure}

\begin{figure}
\includegraphics[width=0.7\textwidth]{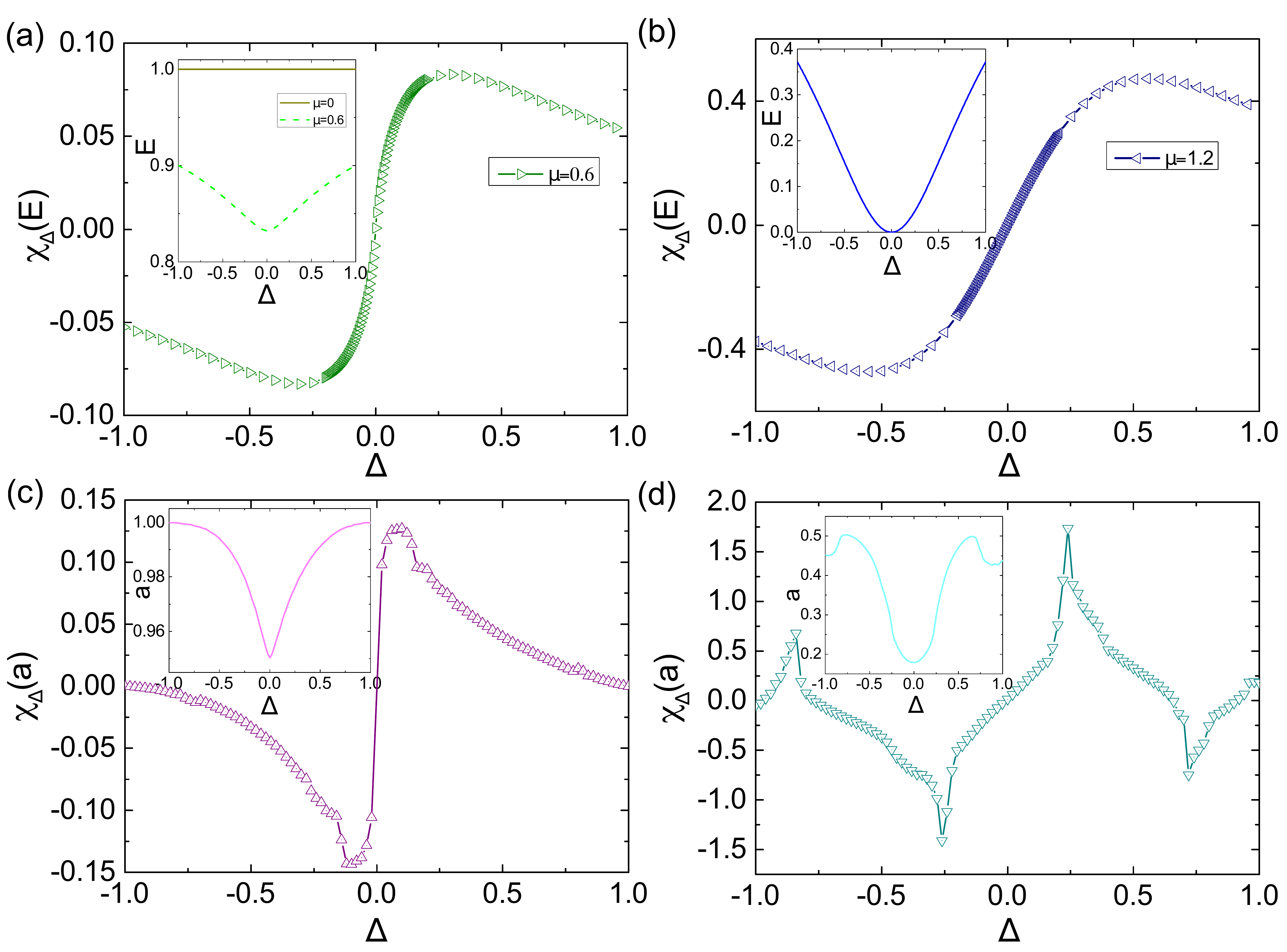}
\caption{(a) The susceptibility of GE $\chi_{\Delta}(E)$ as a function of $\Delta$ in $\mathcal{H}_{1}$ with $\alpha=+\infty$ and $\mu=0.6$. The inset shows the GE $E$ as a function of $\Delta$ in $\mathcal{H}_{1}$ with $\alpha=+\infty$ and $\mu=0,0.6$. (b) The susceptibility of GE $\chi_{\Delta}(E)$ as a function of $\Delta$ in $\mathcal{H}_{1}$ with $\alpha=+\infty$ and $\mu=1.2$. The inset shows the GE $E$ as a function of $\Delta$ in $\mathcal{H}_{1}$ with $\alpha=+\infty$ and $\mu=1.2$. (c) The susceptibility of DE density $\chi_{\Delta}(a)$ as a function of $\Delta$ in $\mathcal{H}_{1}$ with $\alpha=+\infty$ and $\mu=0$. The inset shows the DE density $a$ as a function of $\Delta$ in $\mathcal{H}_{1}$ with $\alpha=+\infty$ and $\mu=0$. (d) Similar to (c) but with $\mu=1.2$.}\label{Fig_a2}
\end{figure}

We study the GE in the extended Kitaev chains to demonstrate its capability in detecting TPTs and compare its results with DE. Firstly, we calculate the GE for the ground states in $\mathcal{H}_{1}$ with $\alpha=0$ and $\mu=0,1.5$, and in this cases, the critical point is $\Delta_{c}=0$ (see Fig. \ref{Fig2}(h)). In Fig. \ref{Fig_a4}, the DE density $a$, the GE $E$ and their susceptibility $\chi_{\Delta}(a)$, $\chi_{\Delta}(E)$, as a function of $\Delta$ are depicted. There are non-analytical behaviors of $a$ and $E$ at the critical point $\Delta_{c}=0$ with $\mu=1.5$. Whereas, only the non-analytical behavior of $a$ at $\Delta_{c}=0$ can be observed for $\mu=0$, while $\chi_{\Delta}(E)$ is continuous at $\Delta_{c}=0$ indicating that GE exhibits analytical behavior. These results show that GE can not efficiently detect TPTs.

We then provide another example to show the instability of GE and the robustness of DE as a probe of TPTs. As shown in Fig. \ref{Fig2}(b) and \ref{Fig2}(f), $\Delta_{c}=0$ is the critical point of TPT in $\mathcal{H}_{1}$ with $\alpha=+\infty$ and $\mu \in (-1,1)$, while for $\mu > 1$ or $\mu < -1$, there is no critical point of TPT. The GE in $\mathcal{H}_{1}$ with $\alpha=+\infty$ and $\mu=0,1.2$ are calculated. In the inset of Fig. \ref{Fig_a2}(a), we find that the GE $E\equiv1$ with $\mu=0$ and different values of $\Delta$, which suggests GE fails to diagnose the TPT. It is also shown that there is a dip of GE at the critical point $\Delta=1$ with $\mu=0.6$. Nevertheless, in Fig. \ref{Fig_a2}(b), similar behavior of GE is also observed at $\Delta=0$ with $\mu=1.2$ where the TPT is absent, and both of the susceptibility $\chi_{\Delta}(E)=\partial E/\partial\Delta$ at $\Delta=0$ with $\mu=0.6$ and $\mu=1.2$ are continuous.

Moreover, the DE density $a$ with the same parameters are also calculated for reference. The results are depicted in Fig. \ref{Fig_a2}(c) ($\mu=0$) and \ref{Fig_a2}(d) ($\mu=1.2$). There is a non-analytical behavior of $a$ at the critical point $\Delta_{c}=0$ with $\mu=0$. But only the local minimum point of $a$ is observed at $\Delta=0$ with $\mu=1.2$, which indicates the absence of TPTs. A more obvious signature can be shown by the susceptibility of $a$ with respect of $\Delta$, i.e., $\chi_{\Delta}(a)=\partial a/\partial\Delta$. The $\chi_{\Delta}(a)=\partial a/\partial\Delta$ at $\Delta=0$ is discontinuous in $\mu=0$, while it is continuous in $\mu=1.2$.

Based on above discussions, the GE can not be a genuine probe of TPTs. In condition that GE fails to characterize TPTs, the DE can still efficiently detect TPTs.

\section{Conclusions}
In conclusion, we have studied the diagonal entropy (DE) in the extended Kitaev chains with variable-range pairing and hopping. Firstly, we show that there is a volume term in the finite size scaling law and block scaling law of DE.
The universality of the volume law is also demonstrated for the topological phases with high winding numbers. Secondly, the DE density, as the parameter of volume term, is numerical calculated. The non-analytical behaviors of DE density, equivalent to the discontinuity of its susceptibility, efficiently spotlight the critical points related to topological phase transitions (TPTs) in the Kitaev chains. Finally, we reveal that
the global entanglement can not explicitly diagnose the TPTs, which makes a merit of DE as a probe of TPTs.

It is well-known that experimentally addressing EE is still a challenge~\cite{exp_a1}. We emphasize that the DE can be measured more easily than EE via quantum tomography since only $2^{N}$ instead of $2^{2N}$ measurements are required for $N$ qubits system~\cite{exp_a2}. Consequently, DE is an experimentally feasible quantity that can detect TPTs.
This work may shed light on a comprehensive understanding of the coherence and DE in topological superconductors. The method is well worthy extending to the characterization of the TPTs in two-dimensional systems~\cite{toric1,toric2,toric3,2D_add} and non-Hermite systems~\cite{non_H1,non_H2,non_H3}, as well as the TPTs with finite temperature~\cite{tem}.

\begin{acknowledgments}
Z.H.S. would like to thank Yu-Ran Zhang for useful discussions. Q.H. was partially supported by the National Key R$\&$D Program of China (Grants No. 2018YFB1107200 and No. 2016YFA0301302),
National Natural Science Foundation of China  (Grants No. 11622428, and No. 61475006), and thanks C.P. Sun at Beijing Computational Science Research Center
for his hospitality. H.F. was partially supported by the National Key R$\&$D Program of China (Grants No. 2016YFA0302104 and No. 2016YFA0300600), National Natural Science Foundation of China (Grant No. 11774406), and Strategic Priority Research Program of Chinese Academy of Science (Grant No. XDB28000000).

\end{acknowledgments}

\appendix
\section{Exact solutions of extended Kitaev chains}
In this section, we briefly review the Hamiltonian, winding numbers and phase diagrams of the extended Kitaev chains with extensive pairing
and hopping terms. We also present additional results of the diagonal entropy in momentum space.

The Hamiltonian of the extended Kitaev chain with variable-range pairing can be written as
\begin{eqnarray}
\mathcal{H}_{1} = -\frac{J}{2}\sum_{j=1}^{N}(c_{j}^{\dagger}c_{j+1}+\text{H.c.})-\mu\sum_{j=1}^{N}(c_{j}^{\dagger}c_{j}-\frac{1}{2})
+ \frac{\Delta}{2}\sum_{j=1}^{N}\sum_{l=1}^{N-j}\frac{1}{d_{l}^{\alpha}}(c_{j}c_{j+l}+\text{H.c.}),
\label{hamiltonian1}
\end{eqnarray}
The parameters have been explained after Eq.~(1) in main text. By applying the Fourier transform:
\begin{eqnarray}
c_{j}=\frac{1}{\sqrt{N}}\sum_{k}e^{ijk}c_{k},\ c_{j}^{\dagger} =\frac{1}{\sqrt{N}}\sum_{k}e^{-ijk}c_{k}^{\dagger},\label{FT}
\end{eqnarray}
the Hamiltonian can be rewritten as
\begin{eqnarray}
\mathcal{H}_{1} = &-&\sum_{k}\frac{1}{2}(J\cos k +\mu)(c_{k}^{\dagger}c_{k}+c_{-k}^{\dagger}c_{-k}) + i\frac{\Delta}{4}\sum_{k}f_{\alpha}(k)(c_{k}c_{-k}-c_{-k}^{\dagger}c_{k}^{\dagger}) \nonumber \\
&=& \sum_{k}\epsilon_{k}\mathbf{\Psi}_{k}^{\dagger}(\mathbf{h}_{k}\cdot\overrightarrow{\sigma})\mathbf\Psi_{k},
\label{hamiltonian1k}
\end{eqnarray}
where $\epsilon_{k} = \pm\sqrt{(J\cos k + \mu)^{2}+(f_{\alpha}(k)\Delta/2)^{2}}$ is the energy spectra, $f_{\alpha}(k)= \sum_{l=1}^{N-1}\sin(kl)/d_{l}^{\alpha}$, ${\mathbf\Psi_{k}^{\dagger}}=(c_{k}^{\dagger},c_{k})$ is the Nambu spinor, $k_{n}=\frac{2\pi}{N}(n+\frac{1}{2})$ $(n=0,1,2,...,N)$ with anti periodic boundary condition, and
\begin{equation}
\mathbf{h}_{k}\cdot\overrightarrow{\sigma} = h_{y}(k)\sigma_{y}+h_{z}(k)\sigma_{z},
\label{hk}
\end{equation}
with
\begin{equation}
h_{y}(k) = - \frac{f_{\alpha}(k)\Delta/2}{\epsilon_{k}}, \ h_{z}(k) = - \frac{J\cos k + \mu}{\epsilon_{k}}. \label{hxy}
\end{equation}
Then the expression of Bogoliubov angle is given by
\begin{equation}
\Theta_{k} = \frac{1}{2}\arctan(\frac{f_{\alpha}(k)\Delta/2}{J\cos k + \mu}). \label{angle}
\end{equation}

\begin{figure*}[]
	\centering
	\includegraphics[width=0.6\linewidth]{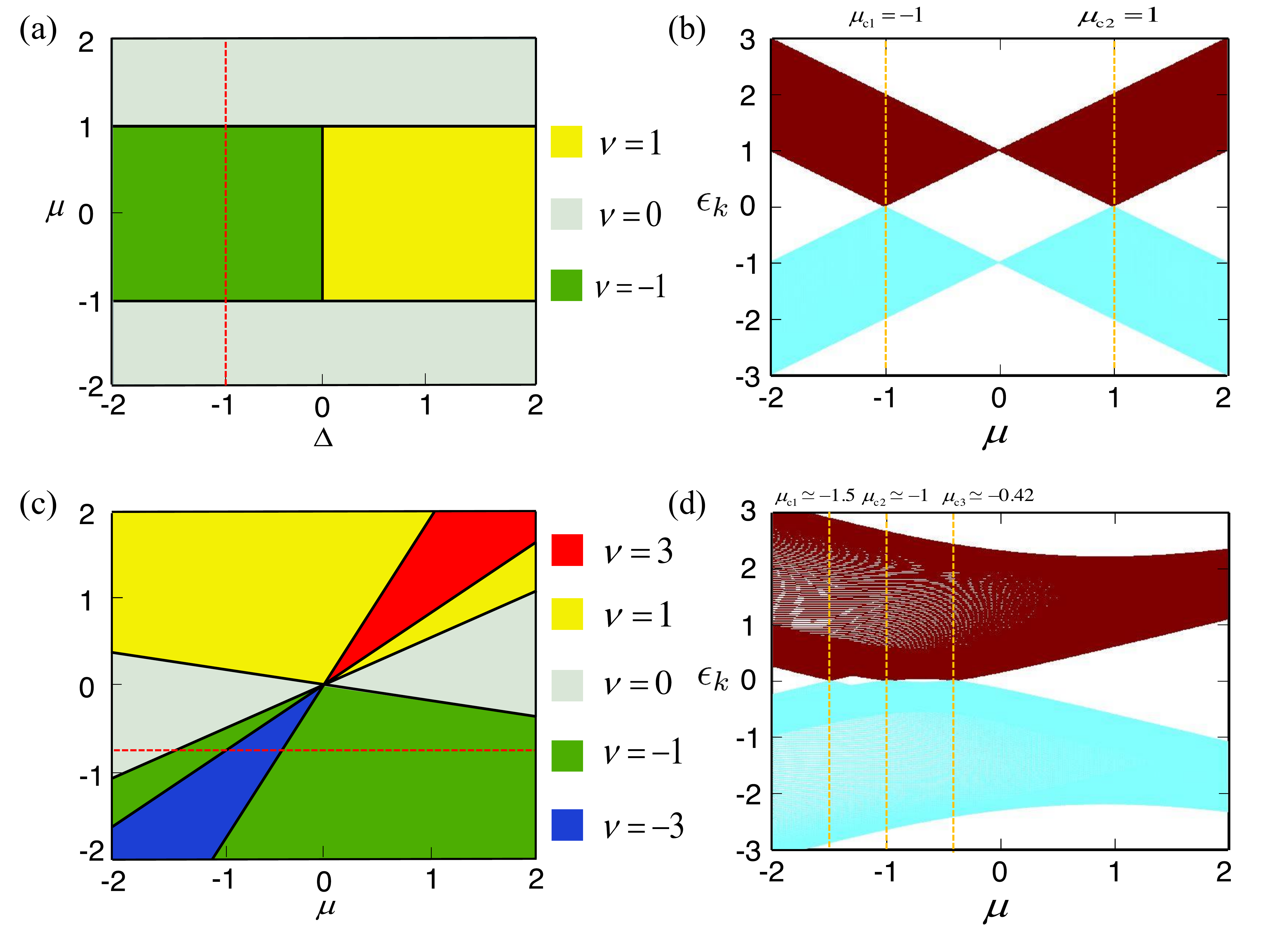}
	\caption{For $\mathcal{H}_{1}$, (a) The winding number $\nu$ in the $\Delta-\mu$ plane when $J=1$, $\alpha=+\infty$. (b) Energy spectra of $\mathcal{H}_{1}$ with $\Delta=-1$ for $N=500$ sites. For $\mathcal{H}_{2}$, (c) The winding number $\nu$ in the $\mu-J$ plane with $r=3$, $\alpha = \beta=0.2$, and $\Delta=1$. (d) Energy spectra of $\mathcal{H}_{2}$ with $J=-0.8$ for $N=500$ sites.}
	\label{Fig5}
\end{figure*}

Because of the $\mathbb{Z}$ symmetry, the topological phases in this model can be characterized by winding numbers
\begin{equation}
\nu = \frac{1}{2\pi}\oint d \Theta = \frac{1}{2\pi}\int_{-\pi}^{\pi}dk \frac{1}{h_{y}(k)}\frac{\partial h_{z}(k)}{\partial k}. \label{wn}
\end{equation}
Moreover, the winding numbers can be directly obtained from the trajectory of the winding vector $\mathbf{h}_{k}=(0,h_{y}(k),h_{z}(k))$.
For instance, when $J=1$, $\alpha=+\infty$, the winding numbers in the $\mu-\Delta$ plane are displayed in Fig.~\ref{Fig5}(a). In addition, the energy spectra can reveal the locations of the critical points related to topological phase transitions (TPTs), see Fig.~\ref{Fig5}(c).

Similarly, the Hamiltonian of the Kitaev chain with longer-range hopping and pairing $\mathcal{H}_{2}$ given in Eq.~(2) in the main text can be rewritten as
\begin{equation}
\mathcal{H}_{2}=\sum_{k}\epsilon_{k}\mathbf\Psi_{k}^{\dagger}(\mathbf{h}_{k}\cdot\overrightarrow{\sigma})\mathbf\Psi_{k}, \label{h2k}
\end{equation}
where
\begin{equation}
h_{y}(k) = \frac{\Delta\sum_{l=1}^{r}\sin(kl)d_{l}^{-\beta}}{\epsilon_{k}}, \ \\
h_{z}(k) = - \frac{\mu/2+J\sum_{l=1}^{r}\cos(kl)d_{l}^{-\alpha}}{\epsilon_{k}}, \label{hxy2}
\end{equation}
with
\begin{equation}
\epsilon_{k}=\pm\sqrt{\left[\Delta\sum_{l=1}^{r}\sin(kl)d_{l}^{-\beta}\right]^{2}+\left[\mu/2+J\sum_{l=1}^{r}\cos(kl)d_{l}^{-\alpha}\right]^{2}}.
\end{equation}
The expression of Bogoliubov angle is given by
\begin{equation}
\Theta_{k} = \frac{1}{2}\arctan\left[\frac{\Delta\sum_{l=1}^{r}\sin(kl)d_{l}^{-\beta}}{\mu/2+J\sum_{l=1}^{r}\cos(kl)d_{l}^{-\alpha}}\right]. \label{angle2}
\end{equation}

In Fig.~\ref{Fig5} (c) and (d), we present the phase diagram, and the energy spectra of $\mathcal{H}_{2}$ with parameters $r=3$, $\alpha = \beta=0.2$, and $\Delta=1$, showing the locations of critical points are $\mu_{c}=-1.5,-1,-0.42$.

\section{The block scaling law of diagonal entropy}
In this section, we present the results of the block scaling law of DE in detail. The basis-dependence property of DE is also discussed.

The diagonal entropy (DE) is dependent on the chosen basis. In this section, we demonstrate that the block scaling law of DE can be represented as a volume term plus a logarithm term on the number of spins and a constant term, and can detect the critical points of TPTs.

We calculated DE for reduced density matrix for a block of $L$ contiguous spins in the extend Kitaev chain. The diagonal terms of the reduced diagonal density matrix $\rho^{\text{diag.}}_L$ for the ground state of the extended Kitaev chains can be expanded as
\begin{equation}
\rho^{\text{diag.}}_{L} = \frac{1}{2^{L}}\sum_{\substack{\gamma_{1},..., \gamma_{L},\in\{0,z\}}}\langle\sigma^{\gamma_{1}}_{1}...\sigma^{\gamma_{L}}_{L}\rangle\sigma^{\gamma_{1}}_{1}...\sigma^{\gamma_{L}}_{L}, \label{diagonal_matrix}
\end{equation}
where $\sigma^{z}_{l}$ and $\sigma^{0}_{l}$ $(l=1,...,L)$ are the Pauli matrix and two-dimensional identity matrix, respectively. The correlation functions $\langle\sigma^{\gamma_{1}}_{1}...\sigma^{\gamma_{L}}_{L}\rangle$ can be calculated by applying the transformation $\sigma^{z}_{j}=-A_{j}B_{j}$. The operators $A_{j}$, $B_{j}$ satisfy
\begin{equation}
\langle A_{l}A_{j}\rangle = \delta_{lj}, \ \langle B_{l}B_{j}\rangle = \delta_{lj}, \ \langle A_{l}B_{j}\rangle = \frac{1}{2\pi}\int_{-\pi}^{\pi}e^{iRk}e^{-2i\Theta_{k}}dk, \label{ABT}
\end{equation}
where $R=j-l$ and $\Theta_{k}$ is the Bogoliubov angle.

Then we can obtain the correlation functions by the Wick theorem. For instance, in the case of $L=4$, the correlation function $\langle\sigma^{z}_{1}\sigma^{z}_{2}\sigma^{z}_{3}\sigma^{z}_{4}\rangle$ can be evaluated as
\begin{eqnarray}
\langle\sigma^{z}_{1}\sigma^{z}_{2}\sigma^{z}_{3}\sigma^{z}_{4}\rangle &=& \langle A_{1}B_{1}A_{2}B_{2}A_{3}B_{3}A_{4}B_{4} \rangle \nonumber \\
&=& \langle A_{1}B_{1}\rangle \langle A_{2}B_{2}A_{3}B_{3}A_{4}B_{4} \rangle
+ \langle A_{1}B_{2}\rangle\langle B_{1}A_{2}A_{3}B_{3}A_{4}B_{4} \rangle + \cdots \nonumber \\
&=& \left|
\begin{array}{cccc}
G_{0} & G_{1} & G_{2} & G_{3} \\
G_{-1} & G_{0} & G_{1} & G_{2} \\
G_{-2} & G_{-1} & G_{0} & G_{1} \\
G_{-3} & G_{-2} & G_{-1} & G_{0} \\
\end{array}
\right|,
\label{ABT1}
\end{eqnarray}
where $G_{R}=1/(2\pi)\int_{-\pi}^{\pi}e^{iRk}e^{-2i\Theta_{k}}dk$.

We have already shown in Figs.~\ref{Fig1}(a) and (b) that the formulation of DE block scaling law is
\begin{equation}
S(\rho_{L}^{\text{diag.}}) = aL + b\log_{2}L +c. \label{DE}
\end{equation}
It is worth stressing that the block scaling law of von Neumann entropy is $S(\rho_{L})\propto \log_{2}L$, and it is reasonable that the second term of DE block scaling law has the form of $\log_{2}L$ since DE can be regarded as a combination of quantum coherence $\mathcal{C}(\rho)$ and von Neumann entropy, that is, $S(\rho_{L}^{\text{diag.}})= \mathcal{C}(\rho_{L})+S(\rho_{L})$.

\begin{figure}
\includegraphics[width=0.6\textwidth]{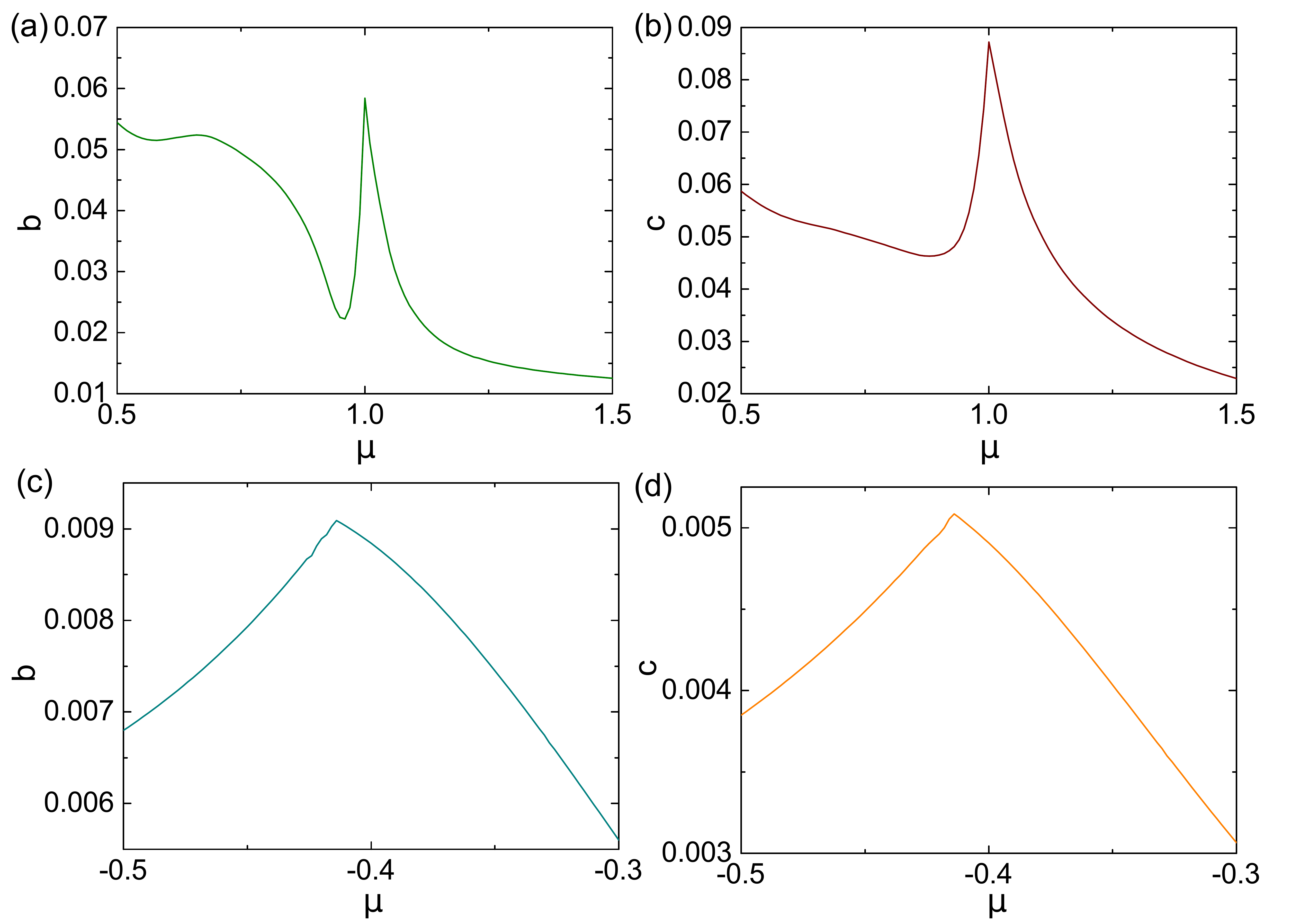}
\caption{(a) The parameter $b$ as a function of $\mu$ for $\mathcal{H}_{1}$ with $\Delta=-1$, $\alpha=0$, $J=1$. The location of critical point is $\mu_{c}=1$. (b) The parameter $c$ as a function of $\mu$ for the same case of (a). (c) The parameter $b$ as a function of $\mu$ for $\mathcal{H}_{2}$ with $\Delta=1$, $\alpha=\beta=0.2$, $J=-0.8$. The location of critical point is $\mu_{c}\simeq -0.42$. (d) The parameter $c$ as a function of $\mu$ for the same case of (c).}\label{Fig8}
\end{figure}

The parameter $a$, described as the DE density extracted from block scaling law, can detect the TPTs, which is shown in the Fig. \ref{Fig4} of the main text. In addition, as shown in Fig.~\ref{Fig8}, the parameters $b$ and $c$ in Eq.~(\ref{DE}) can also spotlight the critical point $\mu_{c}$.

\begin{figure}
\includegraphics[width=0.6\textwidth]{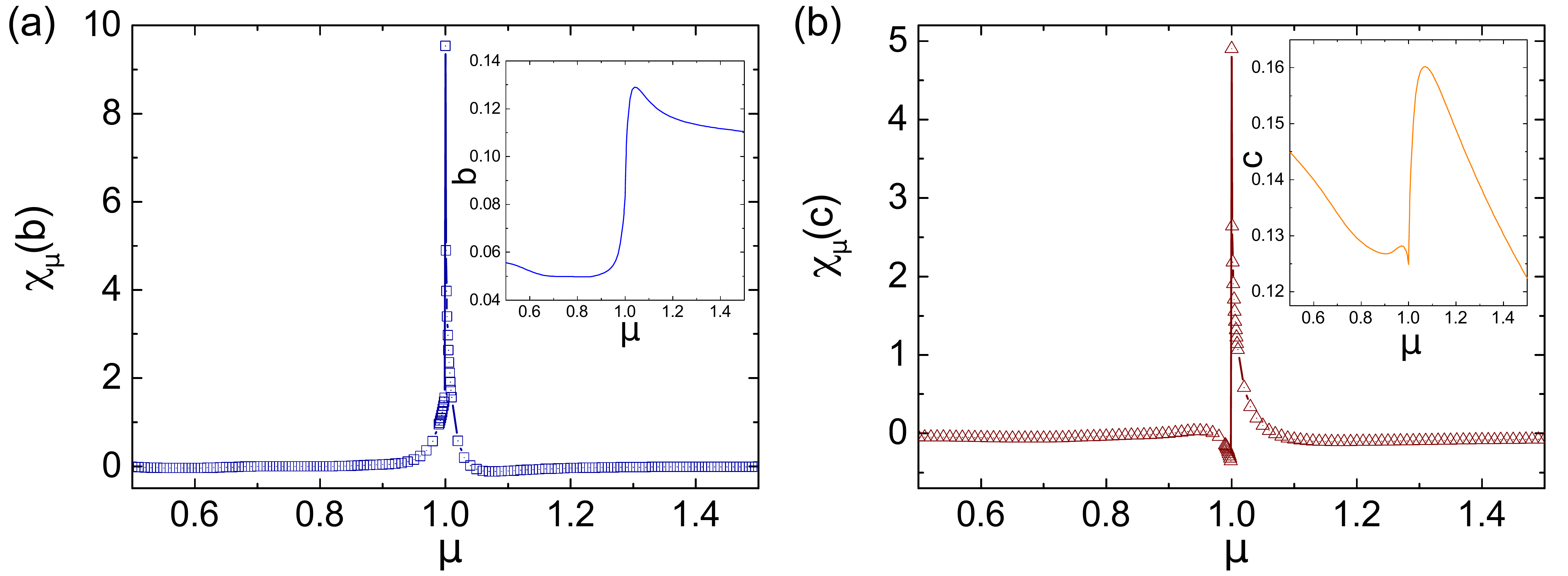}
\caption{(a) The parameter $b$ and its susceptibility with respect of $\mu$ $\chi_{\mu}(b)$ as a function of $\mu$ for $\mathcal{H}_{1}$ with $\Delta=-1$, $\alpha=0$, $J=1$. (b) $c$ and $\chi_{\mu}(c)$ as a function of $\mu$ with the same condition of (a).}\label{Fig9}
\end{figure}

In the $\sigma_{x}$ basis, the the reduced diagonal density matrix $\rho^{\text{diag.}}_L$ can be written as
\begin{equation}
\rho^{\text{diag.}\sigma_{x}}_{L} = \frac{1}{2^{L}}\sum_{\substack{\gamma_{1},..., \gamma_{L},\in\{0,x\}}}\langle\sigma^{\gamma_{1}}_{1}...\sigma^{\gamma_{L}}_{L}\rangle\sigma^{\gamma_{1}}_{1}...\sigma^{\gamma_{L}}_{L},
\label{diagonal_matrix_sigmax}
\end{equation}
where $\sigma_{x}=\left(
\begin{array}{cc}
1 & 0 \\
0 & -1 \\
\end{array}
\right)
$ in its own basis. The correlation functions $\langle\sigma^{\gamma_{1}}_{1}...\sigma^{\gamma_{L}}_{L}\rangle$ $(\gamma_{1},..., \gamma_{L},\in\{0,x\})$ can be calculated by the similar method shown in Eqs.~(\ref{ABT}) and ~(\ref{ABT1}).
The parameters $b$, $c$ and their susceptibility with respect of $\mu$, i.e., $\chi_{\mu}(b)$ and $\chi_{\mu}(c)$ as a function of $\mu$ are shown in Fig.\ref{Fig9}, which indicates that the parameters $b$ and $c$ extracted from the DE in $\sigma_{x}$ basis can also detect the critical points associated with TPTs.
%


\end{document}